\documentclass[apj]{emulateapj}

\usepackage{graphicx, epsfig, amssymb} 
\usepackage{amsmath, amsfonts}
\usepackage{bm} 
\usepackage[linktocpage]{hyperref}
\usepackage[caption=false]{subfig}
\usepackage[usenames]{color}

\def\be{\begin{equation}}
\def\ee{\end{equation}}
\def\beq{\begin{eqnarray}}
\def\eeq{\end{eqnarray}}

\newcommand{\bea}{\begin{eqnarray}}
\newcommand{\eea}{\end{eqnarray}}
\newcommand{\ben}{\begin{enumerate}}
\newcommand{\een}{\end{enumerate}}
\newcommand{\bi}{\begin{itemize}}
\newcommand{\ei}{\end{itemize}}

\newcommand{\nn}{\nonumber}

\slugcomment{Accepted for publication in The Astrophysical Journal.}

\begin{document}

\title{Testing alternative theories of gravity using the Sun}

\author{Jordi Casanellas\altaffilmark{1}, Paolo Pani\altaffilmark{1}, Il\'\i dio Lopes\altaffilmark{1,2}, Vitor Cardoso\altaffilmark{1,3}}

\affil{$^1$CENTRA, Departamento de F\'{\i}sica, 
Instituto Superior T\'ecnico, Universidade T\'ecnica de Lisboa - UTL,
Av.~Rovisco Pais 1, 1049 Lisboa, Portugal.}
\affil{$^2$Departamento de F\'\i sica, Universidade de \'Evora, Col\'egio Luis Ant\'onio Verney, 7002-554 \'Evora - Portugal.} 
\affil{$^3$Department of Physics and Astronomy, The University of Mississippi, University, MS 38677, USA.}

\altaffiltext{}{E-mails: jordicasanellas@ist.utl.pt, paolo.pani@ist.utl.pt, ilidio.lopes@ist.utl.pt, vitor.cardoso@ist.utl.pt}

\begin{abstract}
We propose a new approach to test possible corrections to Newtonian gravity using solar physics. The high accuracy of current solar models and new precise observations allow to constrain corrections to standard gravity at unprecedented levels.
Our case study is Eddington-inspired gravity, an attractive modified theory of gravity which results in non-singular cosmology and collapse. The theory is equivalent to standard gravity in vacuum, but it sensibly differs from it within matter, for instance it affects the evolution and the equilibrium structure of the Sun, giving different core temperature profiles, deviations in the observed acoustic modes and in solar neutrino fluxes. Comparing the predictions from a modified solar model with observations, we constrain the coupling parameter of the theory, $|\kappa_g|\lesssim 3\cdot10^5\text{m}^5\text{s}^{-2}/\text{kg}$. Our results show that the Sun {\it can} be used to efficiently constrain alternative theories of gravity.
\end{abstract}


\maketitle
\date{today}
\section{Introduction}
In the last century General Relativity passed several stringent tests and it is now accepted as the standard theory of gravity and one of mankind's greatest achievements~\citep{Will:2005va}. In the weak-field regime, General Relativity reduces to Newtonian gravity, which is encoded in the famous Poisson equation for the gravitational field
\be
 \nabla^2\Phi=4\pi \mathit{G}\rho\,,\label{Poisson}
\ee
where $\mathit{G}$ is the gravitational constant and $\rho$ is the matter density. In vacuum, the gravitational field of a spherically symmetric mass $M$ simply reads
\be
\Phi(r)=-\mathit{G}M/r\,.\label{inversesquarelaw}
\ee
The validity region of the equation above has been tested and confirmed from submillimeter~\citep{Hoyle:2000cv} to solar system experiments~\citep{Will:2005va}. However, much less is known about Poisson's equation~\eqref{Poisson} inside matter.
In fact, the coupling to matter is one of the most delicate points in Einstein's theory. Several alternative theories have been proposed, which introduce modifications in the coupling between matter and gravity (see e.g. ~\cite{Damour:1993hw}). The investigation of possible alternatives to the General Relativity paradigm are important. Extrapolating Einstein's theory to regimes in which it is not well-tested may lead to bias, potentially affecting astrophysical observations and our understanding of the Universe.

At relativistic level, corrections in the gravity-matter coupling would affect the interior of neutron stars and the cosmological evolution of the universe~\citep{Clifton:2011jh}. However, the uncertainty on the correct equation of state (EOS) describing the interior of a neutron star makes it difficult to disentangle the effects of an alternative theory from those due to a different EOS. 

On the other hand, deviations from standard gravity have been proposed even at Newtonian level~\citep{Milgrom:1983ca,Banados:2010ix} in a way which is compatible with current experimental bounds. Theories such as these are consistent with all observations and at the same time are able to avoid long-standing problems of standard gravity. 
Thus, modified theories should be taken seriously and as important alternatives to explain our Universe and it is of utmost importance to develop methods to test and constrain them against standard gravity.

In this work we propose a new approach, which is not affected by the degeneracy problems in neutron star physics and is complementary to cosmological tests. We shall investigative how deviations in Eq.~\eqref{Poisson} would affect the evolution and the equilibrium structure of the Sun and other stars, leaving potentially observable effects. The high accuracy obtained with current standard solar models and precise observations of the acoustic modes and neutrino fluxes allow to perform stringent tests of the physics governing the star evolution and interior~\citep{TurckChieze:2010gc}. In the past, stellar evolution has been used to constrain a possible time dependence of Newton's constant $\mathit{G}$~\citep{Teller:1948}. More recently, similar ideas have been used to put constraints on the value of $\mathit{G}$~\citep{Lopes:2003aa}, on the properties of dark matter particles~\citep{Lopes:2002gp,Lopes:2010zz,Casanellas:2010he}, and on the couplings of other particles~\citep{Gondolo:2008dd}. Finally, possible modifications to the stellar structure in some alternative scenarios were studied by~\cite{Bertolami:2004nh,Bertolami:2007vu} using polytropic models. Given the high (and increasing) accuracy of present realistic solar models and related observations, using the Sun as a \emph{theoretical laboratory} where alternative theories of gravity can be challenged, is a very promising tool to constrain deviations from Newtonian gravity.

\section{Parametrized Post Poissonian approach for modified gravity}

The Parametrized Post Newtonian approach proved to be extremely efficient to constrain weak-field deviations from General Relativity \emph{in orbital motion} \citep{Will:2005va}. The approach is based on a very general parametrization of the metric functions, and does not require any knowledge of the underlying alternative (metric) theory. Following a similar approach, here we parametrize viable couplings between matter and gravity in the non-relativistic limit, i.e. within Newtonian theory. We require a modified Poisson equation which reduces to the usual one in vacuum, but which can accommodate extra terms in the coupling with matter. Assuming this theory is the non-relativistic limit of some covariant relativistic theory, we also require spacial covariance. Finally, we assume the theory contains at most second-order derivatives in the fields, although this condition can be easily relaxed. A general modified Poisson equation, up to second order in $\Phi$, $\rho$ and derivatives, which satisfies these requirements, reads
\begin{eqnarray}
\nabla^2\Phi&&=4\pi \mathit{G}\rho+\frac{\kappa_g}{4}\nabla^2\rho+\alpha_g\epsilon^{ij}\nabla_i\Phi\nabla_j\rho\nn\\
&&+\eta\rho^2+\gamma\nabla\rho\cdot\nabla\rho +\epsilon_1\nabla\Phi\cdot\nabla\rho \nn\\
&&+\epsilon_2\Phi\nabla^2\rho+\epsilon_3\rho\nabla^2\Phi+...\,\label{P3}
\end{eqnarray}
The first term on the right hand side of the equation above is the standard Poisson term. The second one, proportional to $\kappa_g$, arises from the Eddington-inspired gravity theory recently proposed by \cite{Banados:2010ix}. The other terms are higher order corrections and $\epsilon^{ij}$ is the Levi-Civita symbol. All the parametrized corrections vanish in vacuum, so that the theory above is consistent with the inverse square law behavior~\eqref{inversesquarelaw}, but most of the extra terms in Eq.~\eqref{P3} violate the equivalence principle and are therefore already strongly constrained by experiments~\citep{Will:2005va}. Two notable exceptions are the terms proportional to $\kappa_g$ and, for spherically symmetric configurations, the term proportional to $\alpha_g$. These two terms are consistent with the equivalence principle, and mostly unconstrained presently.

\subsubsection*{A case study}
For concreteness, here we focus on a particular case, setting $\gamma=\eta_i=\epsilon_i=0$ in Eq.~\eqref{P3}. 
The modified Poisson equation reduces to
\begin{equation}
\nabla^2\Phi=4\pi \mathit{G}\rho+\frac{\kappa_g}{4}\nabla^2\rho+\alpha_g\epsilon^{ij}\nabla_i\Phi\nabla_j\rho\,,\label{Poisson2}
\end{equation}
where $[[\kappa_g]]=\text{cm}^5/(\text{g s}^2)=[[\mathit{G}]][[R^2]]$. Requiring spherical symmetry, the hydrostatic equilibrium equation follows
\be
\frac{dP}{dr}=-\frac{\mathit{G}m(r)\rho}{r^2}-\frac{\kappa_g}{4}\rho\rho'\,.\label{hydroeq2}
\ee
where no terms proportional to $\alpha_g$ arise, due to the spherical symmetry.
The choice $\gamma=\eta_i=\epsilon=0$ is motivated by several reasons. First of all, the terms we are neglecting would introduce violations to the equivalence principle, which is experimentally confirmed with great precision~\citep{Will:2005va}. Secondly, the equation above represents the most general modified Poisson equation which is first order in $\Phi$, $\rho$ and satisfies the requirements previously discussed. The extra terms would only introduce higher order corrections. Furthermore, this theory is the nonrelativistic limit of a well-motivated theory of gravity, which prevents the formation of singularities in cosmology and in the stellar collapse of compact objects~\citep{Banados:2010ix,Pani:2011mg}. Here we investigate how this theory would modify the interior and the evolution of the Sun.

\section{The evolution of the Sun}
The high accuracy of current solar models and precise observations allow to test standard gravity against alternative theories at unprecedented levels. Alternative theories would affect the evolution and the equilibrium structure the Sun, giving different core temperature profiles and deviations in the observed acoustic modes and in solar neutrino fluxes. Comparing the predictions from a modified solar model with observations, we can constrain the coupling parameter of the theory.

Modeling the solar interior not only requires to describe the present solar structure, but also to explain the evolution of the Sun from the ignition of hydrogen nuclear fusion to the present day (see e.g.~\cite{TurckChieze:1993dw} for a review). The solar models are constructed on the basis of plausible assumptions, which translate in a set of four ordinary differential equations. The star is considered in hydrostatic equilibrium, which means that the hydrostatic pressure resulting from the thermonuclear fusion of hydrogen to helium must be exactly balanced by gravity. The nuclear reactions are produced in the pp chain and in the CNO cycle, the former affecting more strongly the temperature profile in the solar core.
Furthermore, the star is assumed to be in thermal equilibrium, i.e. the energy produced by nuclear reactions balances the total energy loss via radiative energy flux and via the energy carried away by neutrinos. 

Within $\sim70\%$ of the solar radius, the most efficient transport mechanism of energy from the solar center outwards to the stellar surface is due to electromagnetic radiation, while in the outer region, the so-called convective zone, the energy is mainly transported by convection. 
The radiative energy transport (and, in turn, the temperature profile) is governed by the Rosseland mean opacity, which takes into account that photons interact with electrons and ions in the dense plasma in the solar interior, while they mostly interact with atoms and molecules at the solar surface, where radiative transport is again significant.

One of the basic assumptions of any solar model, namely the hydrostatic equilibrium, ultimately depends on how strong and efficient the gravitational self-interaction is inside the Sun, i.e. on Poisson's equation~\eqref{P3}. Any corrections would affect the thermal balance and, in turn, the temperature profile inside the star, leaving potentially observable signatures.

Finally, effects due to rotation~\citep{Pinso1989ApJ} and magnetic fields~\citep{Passos2008ApJ} are usually neglected in standard solar models. These processes take place on a much shorter timescale than the evolutionary timescale of the Sun and their inclusion results in minor structure changes in the solar interior (see e.g.~\cite{TurckChieze:2010vk}).

\subsection{Equations governing stellar equilibrium and evolution}
Under the previous assumptions, the internal structure of the Sun is governed by the following ordinary differential equations for $(r,P,L,T)$, the radius, pressure, luminosity and temperature, respectively
\begin{eqnarray}
 \frac{dr}{dq}&=&\frac{M_\odot}{4\pi r^2 \rho}\,,\label{mass_conservation}\\
 \frac{dP}{dq}&=&-\frac{\mathit{G} M_\odot^2 q}{4\pi r^4}-\frac{\kappa_g}{4}\rho\frac{d\rho}{dq}\,,\label{hydroeq}\\
 \frac{dL}{dq}&=&M_\odot\left(\epsilon-r\frac{dS}{dt}\right)\,,
\end{eqnarray}
where $q=m/M_\odot$ is a convenient choice of the independent variable, since mass loss is neglected~\citep{Clayton}. The first and third equations above are the standard continuity equation and conservation of thermal energy, respectively, whilst the second equation describes the hydrostatic equilibrium (note that Eq.~\eqref{hydroeq} is equivalent to Eq.~\eqref{hydroeq2} when expressed in terms of independent variable $q$). Finally, the equations above must be supplied by an appropriate transport energy equation, for the convective zone and for the radiative zone~\citep{CESAM}. Due to the modified Poisson equation~\eqref{hydroeq}, the standard equation for the convective energy transport is indirectly modified as follows
  \begin{equation}
  \frac{dT}{dq}\equiv\frac{dP}{dq}\frac{dT}{dP}=-\left[\frac{\mathit{G} M_\odot^2 q}{4\pi r^4}+\frac{\kappa_g}{4}\rho\frac{d\rho}{dq}\right]\frac{T}{P}\nabla\,,\label{convective}
 \end{equation}
where $\nabla\equiv d\log T/d\log P$ is the temperature gradient. For adiabatic changes, the temperature gradient can be simply related to one of the adiabatic exponents, $\nabla_\text{ad}=(\Gamma_2-1)/\Gamma_2$~\citep{2004cgps.book.....W}.
In the radiative zone, the transport energy equation is unaffected by $\kappa_g$ and it simply reads
\begin{equation}
 \frac{dT}{dq}=-\frac{3 M_\odot\kappa}{16\sigma T^3}\frac{L}{16\pi^2 r^4}\,,\label{radiative}
\end{equation}
where $\kappa$ is the Rosseland mean opacity and $\sigma$ is the Boltzmann constant.

\subsection{Numerical procedure}
The modified equations above have been included, together with all the relevant physical processes, in CESAM~\citep{CESAM}, a self-consistent numerical code for stellar structure and evolution. 
The main physical inputs of the solar models are the following: the nuclear reaction rates are taken from \cite{Adelberger:1998qm}, with the~\cite{1977ApJ...212..513M} intermediate screening; the opacities are taken from the OPAL95 tables~\citep{1996ApJ...464..943I} for temperatures above 5600~K and from~\cite{1994ApJ...437..879A} for lower temperatures; we used the tabulated OPAL EOS~\citep{1996ApJ...456..902R}; microscopic diffusion is included following the prescription of~\cite{1993ASPC...40..246M}; finally, the solar abundances are taken from~\cite{2005ASPC..336...25A}. For comparison, all the models were also computed with the older, low-metallicity solar abundances~\citep{Grevesse:1998bj}, leading to nearly identical results. This test stresses the robustness of our approach and shows that our analysis is virtually independent of the current uncertainties of solar modeling.

In order to constrain the values of the coupling parameter $\kappa_g$ which are compatible with present observations of the Sun, we constructed calibrated solar models for different values of $\kappa_g$. The models are calibrated to fit the solar properties with an accuracy of 10$^{-5}$. The calibration is performed by varying the parameters $X_{0}$ (the initial abundance of hydrogen in the young Sun) and $\alpha$ (which parametrizes the efficiency of convection as a mechanism of energy transport), and by fixing the solar surface heavy-element content $(Z/X)_{\odot}=0.0165$, age $t_{\odot}=4.57\;$Gyr, radius R$_{\odot}=6.9599\times10^{10}\;$cm, mass M$_{\odot}=1.9891\times10^{33}\;$g  and luminosity L$_{\odot}=3.846\times10^{33}\;$erg$\;$s$^{-1}$. 

It was possible to construct calibrated solar models for $-0.032 \mathit{G} R_\odot^2\lesssim\kappa_g\lesssim0.02 \mathit{G} R_\odot^2$. For $\kappa_g\lesssim-0.032 \mathit{G} R_\odot^2$, no equilibrium stars can be constructed, in agreement to what is shown in~\cite{Pani:2011mg} for simple polytropic models. On the other hand, for $\kappa_g\gtrsim0.02 \mathit{G} R_\odot^2$ equilibrium stellar configurations can be constructed, but their internal structure is so strongly modified that the observed solar properties (namely $Z/X$, $t_{\odot}$, $R_\odot$, $M_\odot$ and $L_\odot$) cannot be matched simultaneously. In Table~\ref{tab:calibrated_models}, we show the values of $X_0$ and $\alpha$ required to calibrate the solar models, together with the central temperature, density and pressure of the models. 
\begin{table}[!h]
 \begin{tabular}{c | c c | c c c c}
$\kappa_g$ & $X_0$ & $\alpha$ & $T_c$ & $\rho_c$ & $p_c$ \\
$(\mathit{G} R_\odot^2)$ & & & ( 10$^7$ K ) & (g cm$^{-3}$ ) & ( dyn cm$^{-2}$ ) \\ 
\hline
-0.032 & 0.78 & 2.77 & 15.54 & 161.5 & 2.56$\times 10^{17}$ \\
-0.01 & 0.76 & 2.07 & 15.25 & 150.1 & 2.36$\times 10^{17}$ \\
0 & 0.74 & 1.84 &  15.18 & 146.7 & 2.29$\times 10^{17}$ \\
0.01 & 0.73 & 1.65 & 15.12 & 144.8 & 2.25$\times 10^{17}$ \\
0.02 & 0.72 & 1.48 & 15.09 & 143.6 & 2.22$\times 10^{17}$ \\
\end{tabular}
\caption{Characteristics of some solar models for different values of $\kappa_g$. All stellar models have $M=M_{\odot}$, $L=L_{\odot}$, and $R=R_{\odot}$ at the solar age $t_{\odot}=4.57$ Gyr.}
\label{tab:calibrated_models}
\end{table}

Solar models with $\kappa_g>0$ have a lower central density and a lower core temperature, whereas models with $\kappa_g<0$ work in the opposite direction. These results can be qualitatively understood as follows. Equation~\eqref{hydroeq2} can be written in a more evocative form as
\begin{equation}
\frac{dP}{dr}=-\mathit{G}_\text{eff}(r)\frac{m(r)\rho(r)}{r^2}\,,\label{hydroeq_eff}
\end{equation}
where we have defined an ``effective'' Newton's constant
\begin{equation}
 \mathit{G}_\text{eff}(r)\equiv \mathit{G}+\frac{\kappa_g}{4}\frac{r^2\rho'(r)}{m(r)}\,.\label{Geff}
\end{equation}
Since $\rho'(r)<0$ inside the Sun, $\mathit{G}_\text{eff}\lessgtr \mathit{G}$ when $\kappa_g\gtrless0$. When $\kappa_g<0$, we expect a stronger effective gravitational force which, for main sequence stars in hydrostatic equilibrium, leads to an increase in the central temperature and, consequently, in the rate of thermonuclear reactions. The solar models constructed with $\kappa_g<0$ also require a higher initial abundance of hydrogen to match the solar observables. This fact can be explained by homology scaling (see e.g.~\cite{1990sse..book.....K}): the luminosity of a star scales as a high power of $G$ and the mean molecular weight, therefore an effective increase in the gravitational force must be compensated by a decrease in the mean molecular weight (and consequently an increase in the hydrogen abundance) to achieve the same luminosity. This qualitative picture is in agreement with results obtained for different (constant) values of $\mathit{G}$ (cf. Table 1 in~\cite{Lopes:2003aa}).

\section{Results}
\begin{figure}[!t]
  \epsfig{file=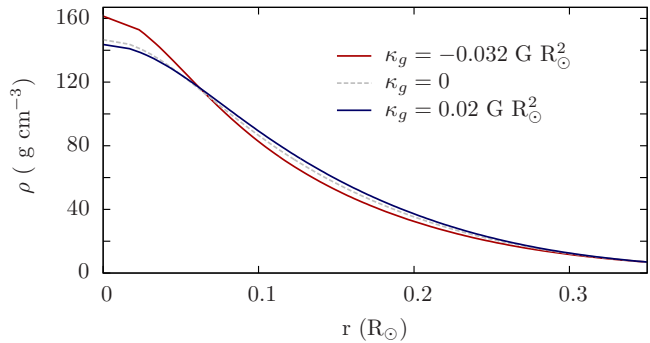,width=8.5cm}
 \caption{Density profiles of the modified solar models computed with different values of $\kappa_g$.}
 \label{fig:rad_profiles}
\end{figure}

The Eddington-inspired theory of gravity leads to strong modifications in the solar structure. In a wide region of the parameter space of the theory, the modified solar models show important variations in the central temperature and in the density profile (see Fig. \ref{fig:rad_profiles}). These two signatures can be tested against solar observables. In particular, we shall show that solar neutrino measurements (which are sensible to $T_c$) and helioseismic acoustic data (sensible to sound speed and density profiles) strongly constrain the values of the parameter $\kappa_g$ that are compatible with present observations.

\subsection{Solar neutrinos}
\label{sec:solar_neutrinos}
Solar neutrinos provide a unique window to the solar interior due to the high sensitivity of thermonuclear reactions to the temperature at which they take place. In particular, the $^8$B flux, produced in the inner 10$\%$ (in radius) through the pp chain, is very sensitive to the central temperature of the Sun: $\phi_{^8B}\propto~T_c^{18}$~\citep{TurckChieze:2010gc}. The predicted neutrino flux is expected to depend strongly on $\kappa_g$ since different couplings lead to different central temperatures (modified solar models may lead to variations of up to 3\% in $T_c$). The observed $^8$B neutrino flux is currently measured with high precision by neutrino telescopes: $(5.046\pm0.16)\times 10^6$ cm$^{-2}$ s$^{-1}$~\citep{Aharmim:2009gd,Bellini:2008mr}. Thus, the theory can be constrained on the basis of incompatibility with observations.

\begin{figure}[!t]
\epsfig{file=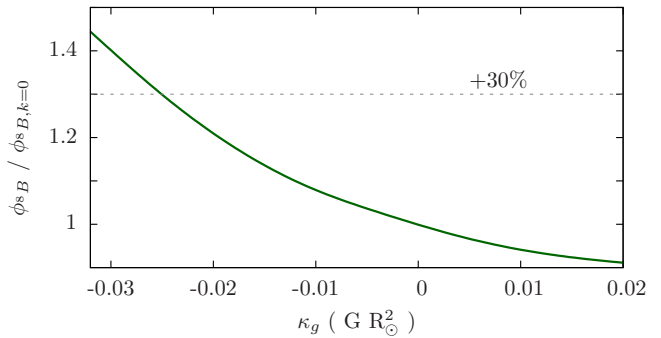,width=8.5cm}
\caption{$^8$B neutrino flux predicted by our modified solar models normalized to the flux predicted by our standard solar model.}
\label{fig:neutrinoflux0}
\end{figure}

Our results for the solar neutrino fluxes are shown in Fig.~\ref{fig:neutrinoflux0}. As expected, the dependence on the coupling parameter $\kappa_g$ can be understood in terms of effective gravitational constant. Positive values of $\kappa_g$ lead to a smaller $\mathit{G}_\text{eff}$, contributing to a lower central temperature and, in turn, to a lower expected neutrino flux. Negative values of $\kappa_g$ work in the opposite direction.

The theoretical uncertainty of standard solar modeling have to be consistently taken into account when comparing the predictions of our models with the observations. Previous works have shown that the largest source of uncertainty in the calculation of the solar neutrino fluxes comes from the uncertainty in the values of the surface heavy element abundances of the Sun~\citep{Bahcall:2005va,2010JHEP...05..072G,Norena:2011sh}. \cite{Bahcall:2004mq} determined, using Monte Carlo simulations for 10000 solar models, that the total 1$\sigma$ theoretical uncertainty in the predicted $^8$B neutrino flux is below 17\%, in the most conservative scenario. In addition, we also take into account the deviation of 20\% in the predicted $^8$B flux when different estimations of the solar abundances are implemented (see e.g.~\cite{Serenelli:2009yc}). Considering both the theoretical and experimental uncertainties, we estimated that models that predict a $^8$B flux which deviates more than 30\% from our standard solar model can be conservatively ruled out, in agreement with the threshold considered by other authors~\citep{Taoso:2010tg}. Following this analysis, we conclude that values of $\kappa_g\lesssim  -0.024 \mathit{G} R_\odot^2$ are excluded by the observation of $^8$B solar neutrinos (see Figure \ref{fig:neutrinoflux0}). 

On the other hand, we found that the $^7$Be neutrino flux only provides less stringent constraints on $\kappa_g$, as $^7$Be neutrinos are produced in a wider region in the center of the Sun and, consequently, are less sensitive to its central temperature ($\phi_{^7\text{Be}}\propto~T_c^{8}$~\citep{TurckChieze:2010gc}).

\subsection{Helioseismology}
\label{sec:helioseismology}
The solar acoustic modes are nowadays measured with exquisite precision by helioseismic missions aboard spacecrafts, such as GOLF/SOHO~\citep{1997SoPh..175..247T}, MDI/SOHO~\citep{1995SoPh..162..129S} and HMI/SDO~\citep{2011SoPh..tmp..163Z}, and by ground networks such as BiSON~\citep{Broomhall:2009up} and GONG~\citep{1996Sci...272.1284H}. The analysis of helioseismic data has provided a valuable tool to probe the solar interior, revealing the sound-speed and density profiles down to 10\% of the solar radius~\citep{ChDals1985Natur,Gough1996Sci}.

Different helioseismic parameters have been used to investigate various aspects of solar physics~\citep{Thompson1996Sci,Gizon2010ARAA}. In particular, the small separation between the frequencies of modes with different degree $l$ and radial order $n$, $\delta\nu_{n,l}=\nu_{n,l}-\nu_{n-1,l+2}$, is a helioseismic quantity which is very sensitive to the temperature gradient in the deep interior of the Sun~\citep{Oti2005MNRAS}. In addition, the modes with degree $l=0$ correspond to acoustic waves that traveled through the entire stellar radius and carry information about the density profile of the Sun~\citep{1994A&A...290..845L,2000MNRAS.317..141R}. Therefore, $\delta\nu_{n,l=0}$, also known as fine spacing or $d_{02}$, is a very suitable parameter to detect the signatures that alternative theories of gravity leave on the solar interior.

\begin{figure}[!t]
\epsfig{file=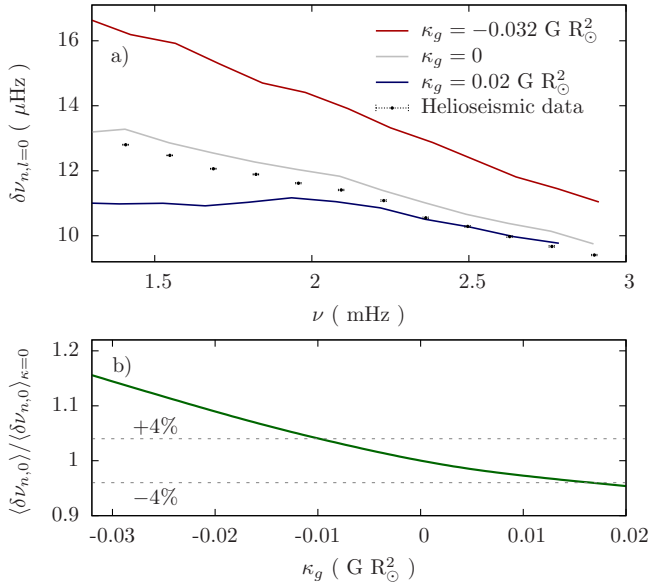,width=8.5cm}
\caption{a) Small separations for l=0 calculated in our models compared with helioseismic data~\citep{Broomhall:2009up}. b) Mean small separation for l=0 and $\nu > 2000 \mu$Hz for our modified solar models, normalized to the prediction for $\kappa_g=0$.}
\label{fig:ssep}
\end{figure}

The small separations in modified solar models are compared with solar data in Figure \ref{fig:ssep}.a). As expected, for $\kappa_g=0$ the fine spacings exhibit a moderate disagreement with the observations. This discrepancy, which disappears when the older, low-Z solar abundances are considered, has been discussed in detail by~\cite{2007ApJ...655..660B}. On the other hand, for large values of $\kappa_g$ the deviations from helioseismic data are much larger, providing a clear way to discriminate viable models.

As discussed for solar neutrinos, when the helioseismic quantities are used to constrain solar models in modified theories of gravity, the uncertainties of solar modeling also have to be taken into account. Compared to solar neutrinos, the theoretical uncertainties on the mean small separation $\langle \delta \nu_{n,l=0} \rangle$ are much smaller. The variation on $\langle \delta \nu_{n,l=0} \rangle$ resulting from different solar models is of the order of 2-3\%~\citep{2007ApJ...655..660B}. Considering this uncertainty, we can rule out those models that lead to deviations in $\langle \delta \nu_{n,l=0} \rangle$ greater than 4\%. This diagnostic establishes strong constraints on $\kappa_g$, ruling out the regions $\kappa_g \gtrsim 0.016$ $\mathit{G}$ R$_\odot^2$ and $\kappa_g \lesssim -0.01$ $\mathit{G}$ R$_\odot^2$. (see Figure \ref{fig:ssep}.b).       

\subsubsection{Other helioseismic constraints}
\begin{figure}[!t]
 \epsfig{file=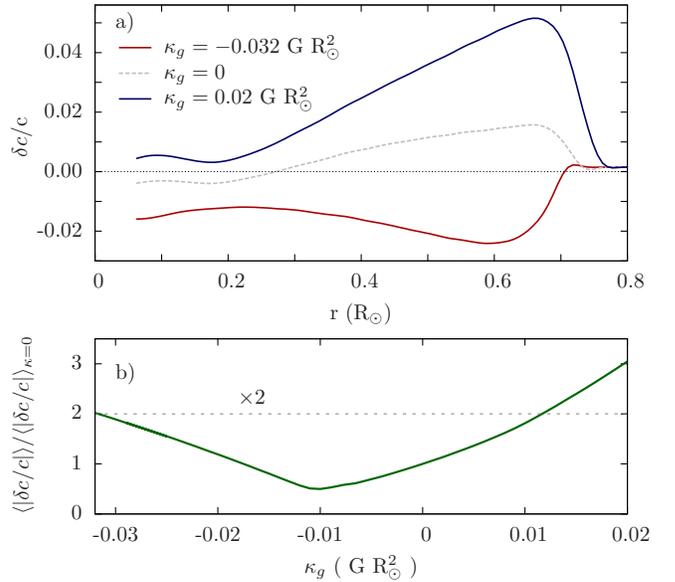,width=8.5cm}
\caption{a) Relative differences between the sound speed profiles of our modified solar models and the solar sound speed from helioseismic data~\citep{Broomhall:2009up}. b) Mean deviation between the solar and model sound speed profiles, normalized to our standard solar model.}
\label{fig:delta-c}
\end{figure}

Another constraint on deviations from Newtonian gravity comes from the comparison of the solar and model sound speed profiles, the former being obtained with high precision from helioseismic observations. Remarkably, the standard solar model reproduces the sound speed profile of the Sun with an accuracy better than 1\% in most of its interior. However, right below the convective envelope the deviations from the observed sound speed are larger (this discrepancy is common in models adopting the latest, high-Z solar abundances~\citep{Montalban:2004za,Delahaye:2005ed,Serenelli:2009yc}). Figure~\ref{fig:delta-c}.a) shows the relative differences between the helioseismically inverted and the sound speed profiles of some of the modified solar models, $\delta c / c = (c_{\odot} - c_{model}) / c_{model}$. The mean difference $\langle | \delta c / c | \rangle$ is a measure of how accurately a solar model reproduces the sound speed profile of the Sun, and therefore it can be used to put constraints to modified theories of gravity. Those models leading to a relative deviation $\langle | \delta c / c | \rangle$ more than two times larger than the $\langle | \delta c / c | \rangle$ obtained for $\kappa_g=0$ can be conservatively ruled out. As shown in Figure \ref{fig:delta-c}.b), the constraints from the sound speed profile rule out models with $\kappa_g\gtrsim0.012$ $\mathit{G}$ R$_\odot^2$.

\begin{figure}[!t]
 \epsfig{file=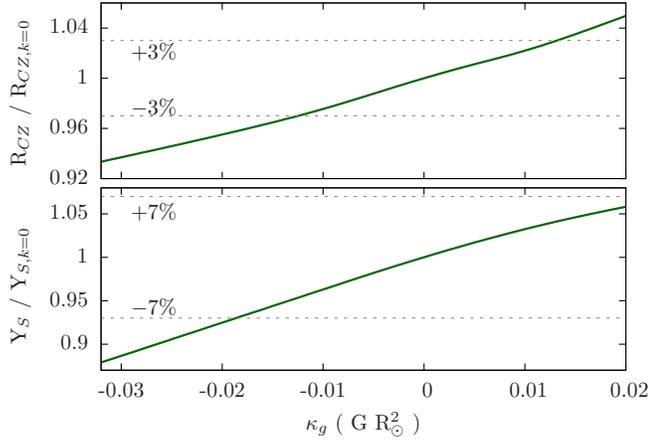,width=8.5cm}
\caption{a) Depth of the convective envelope $R_{CZ}$ and b) helium surface abundance $Y_S$ of the modified solar models normalized to our standard solar model.}
\label{fig:Rcz-Ys}
\end{figure}

Helioseismology also provides accurate measurements of the depth of the convective envelope, $R_{CZ}=0.713\pm0.001$~\citep{BasuAntia:1997} and the helium surface abundance $Y_S=0.2485\pm0.0035$~\citep{BasuAntia:2004}. Monte Carlo simulations have shown that the theoretical uncertainty from solar modeling is below 2\% for $R_{CZ}$ and 5\% for $Y_S$~\citep{Bahcall:2005va}. Consequently, we can conservatively rule out models that predict deviations of these quantities larger than 3\% and 7\%, respectively . As shown in Figure \ref{fig:Rcz-Ys}, this allows us to put the following constraints on the parameter $\kappa_g$: $-0.016 \mathit{G}R_\odot^2<\kappa_g<0.013 \mathit{G} R_\odot^2$ and $\kappa_g > -0.018 \mathit{G} R_\odot^2$, respectively from the observations of $R_{CZ}$ and of $Y_S$.

\section{Discussion and concluding remarks}
Our results show that the Sun is a very good testing ground to constrain generic modified theories of gravity, for instance theories such as the ones described in Eq.~\eqref{P3} and even more exotic or yet to be proposed corrections. For the particular case of Eddington-like theories, Table~\ref{tab:sum_const} summarizes the constraints on the coupling parameter $\kappa_g$ of the theory. Our results show that, in order to obtain a viable solar model, a theory as general as Eq.~\eqref{Poisson2} is strongly constrained.
\begin{table}[!h]
\begin{tabular}{l c}
Observed quantity & Range of $\kappa_g$ excluded \\
\hline
$\phi_{^8B}$ & $\kappa_g < -0.024$ $\mathit{G}$ R$_\odot^2$ \\
$\langle \delta \nu_{n,l=0} \rangle$ &  $\kappa_g < -0.01$ $\mathit{G}$ R$_\odot^2$ and $\kappa_g > 0.016$ $\mathit{G}$ R$_\odot^2$\\
$\langle | \delta c / c | \rangle$ & $\kappa_g > 0.012$ $\mathit{G}$ R$_\odot^2$\\
$R_{CZ}$ & $\kappa_g < -0.013$ $\mathit{G}$ R$_\odot^2$ and $\kappa_g >$ 0.013 $\mathit{G}$ R$_\odot^2$ \\
$Y_S$ & $\kappa_g <$-0.018 $\mathit{G}$ R$_\odot^2$ \\
\end{tabular}
\caption{Summary of the range of the parameter $\kappa_g$ ruled out using different solar characteristics.}
\label{tab:sum_const}
\end{table}
Combining all the experimental bounds, the coupling constant $\kappa_g$ must lie in the interval $-0.01<\kappa_g/(\mathit{G} R_\odot^2)<0.012$, i.e. approximately $|\kappa_g|\lesssim 3\cdot10^5\text{m}^5\text{s}^{-2}/\text{kg}$.

It is important to stress that this result does not rule out Eddington-inspired theory as a promising alternative
to Einstein's theory. Previous studies showed that most of the appealing features of the theory would persist even for a (positive) arbitrarily small coupling parameter~\citep{Banados:2010ix,Pani:2011mg}, which is perfectly consistent with current observations of solar neutrinos and helioseismology.
 
Modified gravity is also relevant as an alternative approach to the solar abundance problem. The particular theory we considered only offers a partial solution to this problem. Indeed, models with $\kappa_g<0$ predict the base of the convective envelope at a smaller radius than the standard solar model, reconciling the prediction with the helioseismically inferred value. However, the predicted helium surface abundance for the same models with $\kappa_g<0$ is then even more underestimated than for standard solar models. Similar partial solutions were discussed in different contexts~\citep{2007A&A...463..755C,2009A&A...494..205C,Guzik:2010ck,Serenelli:2011py}. Although Eddington-inspired gravity suffers from the same limitations, other gravitational corrections could affect the solar interior in a different way and they should be investigated more carefully. We leave this interesting topic for future work.

\begin{acknowledgments}
  We are grateful to the authors of CESAM~\citep{CESAM} and ADIPLS~\citep{ChristensenDalsgaard:2007en} for making their codes publicly available and to Clifford Will for comments and suggestions. This work was supported by the {\it DyBHo--256667} ERC Starting Grant, by FCT - Portugal through PTDC projects FIS/098025/2008, FIS/098032/2008, CTE-AST/098034/2008 and the grant SFRH/BD/44321/2008 and by allocations at cesaraugusta through project AECT-2011-2-0006 and MareNostrum through project AECT-2011-2-0015 at the Barcelona Supercomputing Center (BSC).
\end{acknowledgments}

\bibliographystyle{apj}

\begin{thebibliography}{62}
\expandafter\ifx\csname natexlab\endcsname\relax\def\natexlab#1{#1}\fi

\bibitem[{Adelberger {et~al.}(1998)Adelberger, Austin, Bahcall, Balantekin,
  Bogaert, {et~al.}}]{Adelberger:1998qm}
Adelberger, E.~G., Austin, S.~M., Bahcall, J.~N., Balantekin, A., Bogaert, G.,
  {et~al.} 1998, Rev.Mod.Phys., 70, 1265

\bibitem[{Aharmim {et~al.}(2010)}]{Aharmim:2009gd}
Aharmim, B., {et~al.} 2010, Phys. Rev., C81, 055504

\bibitem[{{Alexander} \& {Ferguson}(1994)}]{1994ApJ...437..879A}
{Alexander}, D.~R., \& {Ferguson}, J.~W. 1994, \apj, 437, 879

\bibitem[{{Asplund} {et~al.}(2005){Asplund}, {Grevesse}, \&
  {Sauval}}]{2005ASPC..336...25A}
{Asplund}, M., {Grevesse}, N., \& {Sauval}, A.~J. 2005, in Astronomical Society
  of the Pacific Conference Series, Vol. 336, Cosmic Abundances as Records of
  Stellar Evolution and Nucleosynthesis, ed. {T.~G.~Barnes III \& F.~N.~Bash},
  25--+

\bibitem[{Bahcall \& Serenelli(2005)}]{Bahcall:2004mq}
Bahcall, J.~N., \& Serenelli, A.~M. 2005, Astrophys. J., 626, 530

\bibitem[{Bahcall {et~al.}(2006)Bahcall, Serenelli, \& Basu}]{Bahcall:2005va}
Bahcall, J.~N., Serenelli, A.~M., \& Basu, S. 2006, Astrophys. J. Suppl., 165,
  400

\bibitem[{Banados \& Ferreira(2010)}]{Banados:2010ix}
Banados, M., \& Ferreira, P.~G. 2010, Phys.Rev.Lett., 105, 011101

\bibitem[{{Basu} \& {Antia}(1997)}]{BasuAntia:1997}
{Basu}, S., \& {Antia}, H.~M. 1997, Mon.Not.Roy.Astron.Soc., 287, 189

\bibitem[{{Basu} \& {Antia}(2004)}]{BasuAntia:2004}
---. 2004, Astrophys.J., 606, L85

\bibitem[{{Basu} {et~al.}(2007){Basu}, {Chaplin}, {Elsworth}, {New},
  {Serenelli}, \& {Verner}}]{2007ApJ...655..660B}
{Basu}, S., {Chaplin}, W.~J., {Elsworth}, Y., {New}, R., {Serenelli}, A.~M., \&
  {Verner}, G.~A. 2007, \apj, 655, 660

\bibitem[{Bellini {et~al.}(2010)}]{Bellini:2008mr}
Bellini, G., {et~al.} 2010, Phys.Rev., D82, 033006

\bibitem[{Bertolami \& Paramos(2005)}]{Bertolami:2004nh}
Bertolami, O., \& Paramos, J. 2005, Phys.Rev., D71, 023521

\bibitem[{Bertolami \& Paramos(2008)}]{Bertolami:2007vu}
---. 2008, Phys.Rev., D77, 084018

\bibitem[{Broomhall {et~al.}(2009)}]{Broomhall:2009up}
Broomhall, A.-M., {et~al.} 2009

\bibitem[{Casanellas \& Lopes(2011)}]{Casanellas:2010he}
Casanellas, J., \& Lopes, I. 2011, Mon. Not. Roy. Astron. Soc., 410, 535

\bibitem[{{Castro} {et~al.}(2007){Castro}, {Vauclair}, \&
  {Richard}}]{2007A&A...463..755C}
{Castro}, M., {Vauclair}, S., \& {Richard}, O. 2007, Astronomy and
  Astrophysics, 463, 755

\bibitem[{Christensen-Dalsgaard(2008)}]{ChristensenDalsgaard:2007en}
Christensen-Dalsgaard, J. 2008, Astrophys. Space Sci., 316, 113

\bibitem[{{Christensen-Dalsgaard} {et~al.}(2009){Christensen-Dalsgaard}, {di
  Mauro}, {Houdek}, \& {Pijpers}}]{2009A&A...494..205C}
{Christensen-Dalsgaard}, J., {di Mauro}, M.~P., {Houdek}, G., \& {Pijpers}, F.
  2009, Astronomy and Astrophysics, 494, 205

\bibitem[{{Christensen-Dalsgaard} {et~al.}(1985){Christensen-Dalsgaard},
  {Duvall}, {Gough}, {Harvey}, \& {Rhodes}}]{ChDals1985Natur}
{Christensen-Dalsgaard}, J., {Duvall}, Jr., T.~L., {Gough}, D.~O., {Harvey},
  J.~W., \& {Rhodes}, Jr., E.~J. 1985, \nat, 315, 378

\bibitem[{{Clayton}(1968)}]{Clayton}
{Clayton}, D.~D. 1968, {Principles of stellar evolution and nucleosynthesis},
  ed. {Clayton, D.~D.}

\bibitem[{Clifton {et~al.}(2011)Clifton, Ferreira, Padilla, \&
  Skordis}]{Clifton:2011jh}
Clifton, T., Ferreira, P.~G., Padilla, A., \& Skordis, C. 2011

\bibitem[{Damour \& Esposito-Farese(1993)}]{Damour:1993hw}
Damour, T., \& Esposito-Farese, G. 1993, Phys.Rev.Lett., 70, 2220

\bibitem[{Delahaye \& Pinsonneault(2006)}]{Delahaye:2005ed}
Delahaye, F., \& Pinsonneault, M. 2006, Astrophys. J., 649, 529

\bibitem[{{Gizon} {et~al.}(2010){Gizon}, {Birch}, \& {Spruit}}]{Gizon2010ARAA}
{Gizon}, L., {Birch}, A.~C., \& {Spruit}, H.~C. 2010, Annual Review of
  Astronomy and Astrophysics, 48, 289

\bibitem[{Gondolo \& Raffelt(2009)}]{Gondolo:2008dd}
Gondolo, P., \& Raffelt, G. 2009, Phys. Rev., D79, 107301

\bibitem[{{Gonzalez-Garcia} {et~al.}(2010){Gonzalez-Garcia}, {Maltoni}, \&
  {Salvado}}]{2010JHEP...05..072G}
{Gonzalez-Garcia}, M.~C., {Maltoni}, M., \& {Salvado}, J. 2010, Journal of High
  Energy Physics, 5, 72

\bibitem[{{Gough} {et~al.}(1996){Gough}, {Kosovichev}, {Toomre}, {Anderson},
  {Antia}, {Basu}, {Chaboyer}, {Chitre}, {Christensen-Dalsgaard},
  {Dziembowski}, {Eff-Darwich}, {Elliott}, {Giles}, {Goode}, {Guzik}, {Harvey},
  {Hill}, {Leibacher}, {Monteiro}, {Richard}, {Sekii}, {Shibahashi}, {Takata},
  {Thompson}, {Vauclair}, \& {Vorontsov}}]{Gough1996Sci}
{Gough}, D.~O., {et~al.} 1996, Science, 272, 1296

\bibitem[{Grevesse \& Sauval(1998)}]{Grevesse:1998bj}
Grevesse, N., \& Sauval, A.~J. 1998, Space Sci. Rev., 85, 161

\bibitem[{Guzik \& Mussack(2010)}]{Guzik:2010ck}
Guzik, J.~A., \& Mussack, K. 2010, Astrophys.J., 713, 1108

\bibitem[{{Harvey} {et~al.}(1996){Harvey}, {Hill}, {Hubbard}, {Kennedy},
  {Leibacher}, {Pintar}, {Gilman}, {Noyes}, {Title}, {Toomre}, {Ulrich},
  {Bhatnagar}, {Kennewell}, {Marquette}, {Patron}, {Saa}, \&
  {Yasukawa}}]{1996Sci...272.1284H}
{Harvey}, J.~W., {et~al.} 1996, Science, 272, 1284

\bibitem[{Hoyle {et~al.}(2001)Hoyle, Schmidt, Heckel, Adelberger, Gundlach,
  {et~al.}}]{Hoyle:2000cv}
Hoyle, C., Schmidt, U., Heckel, B.~R., Adelberger, E., Gundlach, J., {et~al.}
  2001, Phys.Rev.Lett., 86, 1418

\bibitem[{{Iglesias} \& {Rogers}(1996)}]{1996ApJ...464..943I}
{Iglesias}, C.~A., \& {Rogers}, F.~J. 1996, \apj, 464, 943

\bibitem[{{Kippenhahn} \& {Weigert}(1990)}]{1990sse..book.....K}
{Kippenhahn}, R., \& {Weigert}, A. 1990, {Stellar Structure and Evolution}, ed.
  {Kippenhahn, R.~\& Weigert, A.}

\bibitem[{Lopes \& Silk(2010)}]{Lopes:2010zz}
Lopes, I., \& Silk, J. 2010, Science, 330, 462

\bibitem[{{Lopes} \& {Turck-Chi{\`e}ze}(1994)}]{1994A&A...290..845L}
{Lopes}, I., \& {Turck-Chi{\`e}ze}, S. 1994, Astronomy and Astrophysics, 290,
  845

\bibitem[{Lopes {et~al.}(2002)Lopes, Bertone, \& Silk}]{Lopes:2002gp}
Lopes, I.~P., Bertone, G., \& Silk, J. 2002, Mon.Not.Roy.Astron.Soc., 337, 1179

\bibitem[{Lopes \& Silk(2003)}]{Lopes:2003aa}
Lopes, I.~P., \& Silk, J. 2003, Mon.Not.Roy.Astron.Soc., 341, 721

\bibitem[{{Michaud} \& {Proffitt}(1993)}]{1993ASPC...40..246M}
{Michaud}, G., \& {Proffitt}, C.~R. 1993, in Astronomical Society of the
  Pacific Conference Series, Vol.~40, IAU Colloq. 137: Inside the Stars, ed.
  {W.~W.~Weiss \& A.~Baglin}, 246--259

\bibitem[{Milgrom(1983)}]{Milgrom:1983ca}
Milgrom, M. 1983, Astrophys.J., 270, 365

\bibitem[{{Mitler}(1977)}]{1977ApJ...212..513M}
{Mitler}, H.~E. 1977, \apj, 212, 513

\bibitem[{{Montalb{\'a}n} {et~al.}(2004){Montalb{\'a}n}, {Miglio}, {Noels},
  {Grevesse}, \& {di Mauro}}]{Montalban:2004za}
{Montalb{\'a}n}, J., {Miglio}, A., {Noels}, A., {Grevesse}, N., \& {di Mauro},
  M.~P. 2004, in ESA Special Publication, Vol. 559, SOHO 14 Helio- and
  Asteroseismology: Towards a Golden Future, ed. {D.~Danesy}, 574--+

\bibitem[{{Morel}(1997)}]{CESAM}
{Morel}, P. 1997, A \& A Supplement series, 124, 597

\bibitem[{Nore\~na {et~al.}(2011)Nore\~na, Verde, Jimenez, Pe\~na Garay, \&
  Gomez}]{Norena:2011sh}
Nore\~na, J., Verde, L., Jimenez, R., Pe\~na Garay, C., \& Gomez, C. 2011

\bibitem[{{Ot{\'{\i}} Floranes} {et~al.}(2005){Ot{\'{\i}} Floranes},
  {Christensen-Dalsgaard}, \& {Thompson}}]{Oti2005MNRAS}
{Ot{\'{\i}} Floranes}, H., {Christensen-Dalsgaard}, J., \& {Thompson}, M.~J.
  2005, MNRAS, 356, 671

\bibitem[{Pani {et~al.}(2011)Pani, Cardoso, \& Delsate}]{Pani:2011mg}
Pani, P., Cardoso, V., \& Delsate, T. 2011, Phys.Rev.Lett., 107, 031101

\bibitem[{{Passos} \& {Lopes}(2008)}]{Passos2008ApJ}
{Passos}, D., \& {Lopes}, I. 2008, \apj, 686, 1420

\bibitem[{{Pinsonneault} {et~al.}(1989){Pinsonneault}, {Kawaler}, {Sofia}, \&
  {Demarque}}]{Pinso1989ApJ}
{Pinsonneault}, M.~H., {Kawaler}, S.~D., {Sofia}, S., \& {Demarque}, P. 1989,
  \apj, 338, 424

\bibitem[{{Rogers} {et~al.}(1996){Rogers}, {Swenson}, \&
  {Iglesias}}]{1996ApJ...456..902R}
{Rogers}, F.~J., {Swenson}, F.~J., \& {Iglesias}, C.~A. 1996, \apj, 456, 902

\bibitem[{{Roxburgh} \& {Vorontsov}(2000)}]{2000MNRAS.317..141R}
{Roxburgh}, I.~W., \& {Vorontsov}, S.~V. 2000, Mon.Not.Roy.Astron.Soc., 317,
  141

\bibitem[{{Scherrer} {et~al.}(1995){Scherrer}, {Bogart}, {Bush}, {Hoeksema},
  {Kosovichev}, {Schou}, {Rosenberg}, {Springer}, {Tarbell}, {Title},
  {Wolfson}, {Zayer}, \& {MDI Engineering Team}}]{1995SoPh..162..129S}
{Scherrer}, P.~H., {et~al.} 1995, \solphys, 162, 129

\bibitem[{Serenelli {et~al.}(2009)Serenelli, Basu, Ferguson, \&
  Asplund}]{Serenelli:2009yc}
Serenelli, A., Basu, S., Ferguson, J.~W., \& Asplund, M. 2009, Astrophys. J.,
  705, L123

\bibitem[{Serenelli {et~al.}(2011)Serenelli, Haxton, \&
  Pena-Garay}]{Serenelli:2011py}
Serenelli, A.~M., Haxton, W.~C., \& Pena-Garay, C. 2011

\bibitem[{Taoso {et~al.}(2010)Taoso, Iocco, Meynet, Bertone, \&
  Eggenberger}]{Taoso:2010tg}
Taoso, M., Iocco, F., Meynet, G., Bertone, G., \& Eggenberger, P. 2010, Phys.
  Rev., D82, 083509

\bibitem[{Teller(1948)}]{Teller:1948}
Teller, E. 1948, Phys. Rev., 73, 801

\bibitem[{{Thompson} {et~al.}(1996){Thompson}, {Toomre}, {Anderson}, {Antia},
  {Berthomieu}, {Burtonclay}, {Chitre}, {Christensen-Dalsgaard}, {Corbard}, {De
  Rosa}, {Genovese}, {Gough}, {Haber}, {Harvey}, {Hill}, {Howe}, {Korzennik},
  {Kosovichev}, {Leibacher}, {Pijpers}, {Provost}, {Rhodes}, {Schou}, {Sekii},
  {Stark}, \& {Wilson}}]{Thompson1996Sci}
{Thompson}, M.~J., {et~al.} 1996, Science, 272, 1300

\bibitem[{Turck-Chi{\`e}ze \& Couvidat(2011)}]{TurckChieze:2010gc}
Turck-Chi{\`e}ze, S., \& Couvidat, S. 2011, Rept.Prog.Phys., 74, 086901

\bibitem[{Turck-Chi{\`e}ze \& Lopes(1993)}]{TurckChieze:1993dw}
Turck-Chi{\`e}ze, S., \& Lopes, I. 1993, Astrophys.J., 408, 347

\bibitem[{Turck-Chi{\`e}ze {et~al.}(2010)Turck-Chi{\`e}ze, Palacios, Marques,
  \& Nghiem}]{TurckChieze:2010vk}
Turck-Chi{\`e}ze, S., Palacios, A., Marques, J.~P., \& Nghiem, P. A.~P. 2010,
  Astrophys. J., 715, 1539

\bibitem[{{Turck-Chi{\`e}ze} {et~al.}(1997){Turck-Chi{\`e}ze}, {Basu}, {Brun},
  {Christensen-Dalsgaard}, {Eff-Darwich}, {Lopes}, {P{\'e}rez Hern{\'a}ndez},
  {Berthomieu}, {Provost}, {Ulrich}, {Baudin}, {Boumier}, {Charra}, {Gabriel},
  {Garcia}, {Grec}, {Renaud}, {Robillot}, \& {Roca
  Cort{\'e}s}}]{1997SoPh..175..247T}
{Turck-Chi{\`e}ze}, S., {et~al.} 1997, Solar Physics, 175, 247

\bibitem[{{Weiss} {et~al.}(2004){Weiss}, {Hillebrandt}, {Thomas}, \&
  {Ritter}}]{2004cgps.book.....W}
{Weiss}, A., {Hillebrandt}, W., {Thomas}, H.-C., \& {Ritter}, H. 2004, {Cox and
  Giuli's Principles of Stellar Structure}, ed. {Weiss, A., Hillebrandt, W.,
  Thomas, H.-C., \& Ritter, H.}

\bibitem[{Will(2005)}]{Will:2005va}
Will, C.~M. 2005, Living Rev.Rel., 9, 3, an update of the Living Review article
  originally published in 2001

\bibitem[{{Zhao} {et~al.}(2011){Zhao}, {Couvidat}, {Bogart}, {Parchevsky},
  {Birch}, {Duvall}, {Beck}, {Kosovichev}, \& {Scherrer}}]{2011SoPh..tmp..163Z}
{Zhao}, J., {et~al.} 2011, \solphys, 163

\end{thebibliography}

\end{document}